\documentclass[aps,prl,twocolumn,groupedaddress]{revtex4}
\usepackage{graphicx}
\usepackage{amsmath,amsfonts,amssymb}
\usepackage{color}
\begin{document}
\title{Schwarzschild Mass Uncertainty}

\author{Aharon Davidson}
\email{davidson@bgu.ac.il}
\homepage[Homepage: ]{http://www.bgu.ac.il/~davidson}
\author{Ben Yellin}
\email{yellinb@bgu.ac.il}

\affiliation{Physics Department, Ben-Gurion University of the Negev,
Beer-Sheva 84105, Israel}

\date{September 9, 2012}

\begin{abstract}
Applying Dirac's procedure to $r$-dependent constrained
systems, we derive a reduced total Hamiltonian, resembling
an upside down harmonic oscillator, which generates the
Schwarzschild solution in the mini super-spacetime.
Associated with the now $r$-dependent Schrodinger equation
is a tower of localized Guth-Pi-Barton wave packets, orthonormal
and non-singular, admitting equally spaced average-'energy' levels.
Our approach is characterized by a universal quantum mechanical
uncertainty structure which enters the game already at the flat
spacetime level, and accompanies the massive Schwarzschild
sector for any arbitrary mean mass.
The average black hole horizon surface area is linearly quantized.
\end{abstract}

\pacs{04.70.Dy, 04.60.Kz}

\maketitle

In the beginning Newton postulated a universal gravitational force
law exerted by a massive point particle on bodies floating in a flat
background space and sharing an invariant ticking time.
Special relativity, while unifying space and time, has not challenged
the flatness of the resulting spacetime nor its non-dynamical role.
The general relativity revolution elevated spacetime into  a dynamical
object, albeit classical, with Newton's law embedded within the
celebrated yet singular Schwarzschild solution.
The next challenge is to reveal how would the concept of spacetime
be revised when quantum mechanics enters the game, and how would
Newton's force law, and in particular the black hole event horizon
fit in.
Given the fact that quantum gravity, with its anticipated Planck
scale effects, is still absent, the conventional assumption is that
probing the structure of spacetime is premature at this stage.
In this respect, however, the practical lesson to be deduced from
Hawking-Bekenstein \cite{BH} black hole thermodynamics is
that some combined gravitational and quantum mechanical
effects may be currently attainable after all.
In this paper, generalizing Dirac's procedure for circumferential
radius dependent constrained systems, we quantize the 
Schwarzschild black hole solution (actually the static spherically
symmetric geometry). 
We do it  within the framework of the mini superspace \cite{mini},
differing however from the Kuchar approach \cite{Kuchar},
with the aspiration, however, to shed some light on the quantum
mechanical structure of spacetime in general.
The notion of a source point particle is replaced by a localized
orthonormal Guth-Pi-Barton \cite{GP} wave packet, and
the universal quantum mechanical mass uncertainty structure
which governs the Schwarzschild vacuum/massive black hole
is revealed.

Black hole thermodynamics is anchored to the area entropy
formula \cite{BH}.
The latter points towards some kind of microphysical degrees
of freedom, but does not tell us what they are, where they
live, and how to count them.
In the absence of a satisfactory general relativistic quantum
mechanical answers, it has been argued that the resolution
must lie beyond general relativity, in string theory to be precise.
Truly, stringy black holes have been proven valuable in
this respect, supporting the holographic principle \cite{Hprinciple},
and providing at least a partial answer \cite{stringy} in terms
of D-branes.
They shed light on the general issue, on extremal black hole
in particular, but unfortunately, not directly on the (say)
Schwarzschild black hole.
In this paper, however, we do without relying on string theory
or loop quantum gravity.
Our approach differs both conceptually as well as technically
from previous approaches
\cite{Q}, notably from the
 conventional Kuchar approach \cite{Kuchar}.
For simplicity, we adopt the Planck units $c=G=\hbar=k_B=1$.

Let our starting point be the static spherically symmetric
line element
\begin{equation}
	ds^2=-T(r)dt^2+\frac{dr^2}{R(r)}+S^2(r)d\Omega^2 ~.
\end{equation}
A gauge fixing option is still at our disposal, but as far as
the forthcoming Hamiltonian formalism is concerned, we have
to exercise it with extra caution.
The more so at the mini superspace level, where the general
relativistic action
$\int {\cal R}\sqrt{-g}~d^4 x$ is integrated out over time and
solid angle into $\int {\cal L}(T,R,S,r)dr$.
After integrating out the second derivative surface terms we
are left with
\begin{equation}
	{\cal L}(T,R,S,r)=-\left(1+R {S^{\prime}}^2
	+\frac{R T^{\prime}S S^{\prime} }{T}\right)
	\sqrt{\frac{T}{R}}~,
\end{equation}
where $T(r),R(r),S(r)$ serve as canonical variables.
From this point on, the $r$-evolution of the system is treated
in full analogy with the $t$-evolution in analytical mechanics.

The non-dynamical $R(r)$ is the analogous lapse function.
It is well known that pre-fixing $R(r)$, say $R(r)=1$, is 
problematic.
At the classical level, the algebraic Hamiltonian
constraint which in turn adds a superfluous
degree of freedom to the Schwarzschild solution is gone.
Once $R(r)$ is elevated to the level of an essential canonical
variable, the question is whether the canonical role of $T(r),S(r)$
can be relaxed?
In other words, having in mind canonical quantum gravity
\cite{ADM}, can one harmlessly pre-fix one of them
before conducting the variation?

The answer to this question is in the affirmative, and can be
supported by a simple example.
By pre-fixing the circumferential radius $S(r)=r$,  one can
easily verify that the residual Lagrangian
\begin{equation}
	{\cal L}(T,R,r)=\left(r R^{\prime}+R-1\right)\sqrt{\frac{T}{R}}
	\label{L}
\end{equation}
does produce the exact Schwarzschild solution and nothing else.
In the mini superspace Lagrangian formalism, the coordinate
$r$ can still be redefined $r\rightarrow f(r)$ (involving an
explicit function of $r$), but does not take any part whatsoever
in the variational process itself.
it is thus advantageous, and in some respect even necessary,
to use a pre-gauge which is capable of constituting an invariant
geometrical quantity which is furthermore canonical variable
independent.
For the hereby adopted $S(r)=r$ gauge, it is the invariant
spherical surface area $A(r)=4\pi r^2$ which is
$T,R$-independent as required.
The procedure has been successfully implemented in the
cosmological case \cite{honorable}, where an admissible
gauge choice fixes not the lapse function, but rather the
scale factor.

The path to the quantum black hole is governed by the
Hamiltonian formalism, and to be more specific, by Dirac's
prescription \cite{Dirac} for dealing with constraint systems.
Given the Lagrangian eq.(\ref{L}), and borrowing the language
of analytical mechanics, the corresponding momenta
$\displaystyle{p_R=\frac{\partial {\cal L}}{\partial R^{\prime}}}$
and $\displaystyle{p_T=\frac{\partial {\cal L}}{\partial T^{\prime}}}$
fail to determine the velocities $R^{\prime}$ and $T^{\prime}$.
This, in turn, gives rise to the two primary constraints
\begin{equation}
	\phi_1 = p_R-r\sqrt{\frac{T}{R}}\approx 0 ~, \quad
	\phi_2=p_T\approx 0 ~.
	\label{O}
\end{equation}
The fact that their Poisson brackets does not vanish
\begin{equation}
	\{\phi_1,\phi_2\}=-\frac{r}{2\sqrt{TR}} \neq 0 ~, 
	\label{Poisson}
\end{equation}
makes them second class constraints.
As argued by Dirac, the naive Hamiltonian
${\cal H}_{naive}=p_R R^{\prime}+p_T T^{\prime}-{\cal L}$
is not uniquely determined, and one may add to it any linear
combination of the $\phi$'s, which are zero, and go over to
${\cal H}^{\star}={\cal H}_{naive}+\sum_i u_i \phi_i $.
Consistency then requires the constraints be constants of
motion, and as such, they must weakly obey
\begin{equation}
	\frac{d \phi_i}{dr}=\{\phi_i,{\cal H}_{naive}\}
	+\sum_j u_j \{\phi_i,\phi_j\}
	+\frac{\partial \phi_i}{\partial r} \approx 0~.
	\label{constants}
\end{equation}
Calculating the various Poisson brackets, we
solve these linear equations to find out that
\begin{equation}
	u_R= \frac{1-R}{r} ~, \quad  u_T= \frac{T(1-R)}{rR}  ~. 
\end{equation}
Finally, substituting the coefficients $u_{R,T}$ into ${\cal H}^{\star}$
constitutes the so-called total Hamiltonian
\begin{equation}
	{\cal H}_{total}=(1-R)\left(\sqrt{\frac{T}{R}}
	+\frac{1}{r}\left( p_R -r\sqrt{\frac{T}{R}}\right)
	+\frac{T p_T}{rR } \right)~,	
\end{equation}
which obviously has nothing to do with the ADM Hamiltonian.

A generalization of Dirac's prescription for time dependent
($r$-dependent in our case) constrained Hamiltonians
\cite{tconstraint} is in order.
One is quite familiar with the Dirac brackets technique, 
invoked to make the entire set of constraints first class, 
but in the presence of explicit time dependence (in the
Hamiltonian and/or in the constraints themselves), an extra
step must be taken.
And indeed, the constraints leave their impact on the equations of
motion via the $r$-evolution operator formula
\begin{eqnarray}
	\frac{d }{dr}=[~,{\cal H}_{total}]_D
	+\left.\frac{\partial }{\partial r}\right |_D ~,
\end{eqnarray}
where
\begin{eqnarray}
	& \displaystyle{[X,Y]_D\equiv\{X,Y\}
	+\frac{\epsilon_{ij}}{\{\phi_1,\phi_2\}}
	\{X,\phi_i\}\{\phi_j,Y\}} ~, &\\
	& \displaystyle{\left.\frac{\partial X}{\partial r}\right|_D
	\equiv\frac{\partial X}{\partial r}
	+\frac{\epsilon_{ij}}{\{\phi_1,\phi_2\}}
	\{X,\phi_i\}\frac{\partial \phi_j}{\partial r}} ~. &
\end{eqnarray}
As consistency checks we have verified that all Dirac brackets
involving the $\phi_{1,2}$ constraints vanish, and so do the
dressed partial derivatives
$\displaystyle{\left.\frac{\partial \phi_i}{\partial r}\right|_D}$,
and have reassured the emergence of the classical Schwarzschild
solution.
Among the non-vanishing Dirac brackets we pick up to present
the conventional $\left[R,p_R\right]_D=1$, accompanied by the
unconventional
\begin{equation}
	\left[R,T\right]_D=\frac{2\sqrt{TR}}{r} ~,
	\label{RT}
\end{equation}
which are both of relevance for our forthcoming discussion.
Eq.(\ref{RT}) comes with a message; it simply tells us that the
Schwarzschild metric components $T,R$ would not commute
when elevated to the level of quantum mechanical
operators.

Explicitly imposing now the $\phi_{1,2}$ constraints (thereby
importing them to the quantum level), and subsequently
substituting
\begin{equation}
	T=\frac{p_R R^2}{r^2} ~,
	\label{T}
\end{equation}
we are led to the reduced Hamiltonian
\begin{equation}
	{\cal H}(R,p_R,r)_{reduced}=\frac{1}{r}(1-R)p_R ~,
	\label{reduce}
\end{equation}
subject to the canonical Poisson brackets $\left[R,p_R\right]_P=1$.
As a primary check, one can straight forwardly confirm
the Schwarzschild solution $R=1-2m/r,~p_R=\omega r$
(choosing $\omega$ is nothing but rescaling $t$).
The physical role played by the momentum $p_R$ is manifest
via eq.(\ref{T}) which serves now as the connection with the
underlying metric.

A radial marker redefinition $r=e^{\rho}$ then
transforms the $r$-dependent reduced Hamiltonian
eq.(\ref{reduce}) into an $\rho$-independent variant of
the $xp$-type \cite{xp} discussed in the context of 
Riemann zeta function zeroes.
A successive linear canonical transformation
\begin{equation}
	1-R=\frac{1}{\sqrt{2}}(p-x)~,  \quad
	p_R=\frac{1}{\sqrt{2}}(p+x)~,
	\label{xp}
\end{equation}
yields the upside-down harmonic oscillator Hamiltonian
\begin{equation}
	{\cal H}=\frac{1}{2}\left(p^2-x^2\right)~.
\end{equation}
%%%%%
The inverted harmonic oscillator was previously discussed
\cite{Brout} in black hole physics in the context of Rindler
observers.
%%%%% 
Combined with the latter Hamiltonian is the $\rho$-dependent
Schrodinger equation
\begin{equation}
	-\frac{\partial^2 \psi}{\partial x^2}-x^2\psi
	=2i\frac{\partial \psi}{\partial \rho}~.
\end{equation}
The 'energy' eigenfunctions, proportional to the Hermite
polynomials
$\psi_E \sim e^{\mp \frac{i x^2}{2}}
	H(-\frac{1}{2} \mp iE, \pm e^{\pm\frac{ i\pi}{4}}x)
	e^{-iE\rho}$~,
pose a major problem.
Owing to their $1/\sqrt{|x|}$ behaviors at
$x\rightarrow \pm \infty$, they are not square integrable.
Counter intuitively, however, especially when dealing with an
unbounded potential, there exists a set of localized
wave packets which satisfy the above $\rho$-dependent
Schrodinger equation.

\medskip
\noindent {\underline{\bf{The massless case}}}

The basic wave packets are of the generic form
\begin{equation}
	\psi_n (x,\rho)=P_n (x,\rho)
	 ~e^{-\frac{x^2}{2~}tan(\varphi-i\rho)}~,
	 \label{Pe}
\end{equation}
where $P_n (x,\rho,\varphi)=\sum_{k=0}^n c_k (\rho,\varphi) x^k$
are even/odd polynomials of order $n$.
Note that the differential equation
\begin{equation}
	\frac{\partial^2 P}{\partial x^2}
	-\tan (\varphi-i\rho)
	\left (P+2x \frac{\partial P}{\partial x}\right)
	+2i\frac{\partial P}{\partial \rho}=0
\end{equation}
does allow for a more general series expansion, namely
$P=c_n x^n+c_{n-2}x^{n-2}+...$, but unless $n$ is an integer,
the series doers not terminate, turning the solution singular.
The wave packet solution eq.(\ref{Pe}) is characterized by a real
parameter $\varphi$ which controls the $\rho$-dependent width
of the wave packet
\begin{equation}
	\delta(\rho, \varphi)= \left(\frac{\cos 2\varphi+\cosh2\rho}
	{2\sin 2\varphi}\right)^{1/2} ~.
\end{equation}
The condition $\sin 2\varphi>0$ then suffices to assure the tenable
behavior $\psi_n \rightarrow 0$ as $x \rightarrow \pm\infty$.
Altogether, being non singular \cite{singular}, square integrable,
and furthermore orthonormal
$\int_{-\infty}^\infty \psi^{\dag}_n \psi_m dx=\delta_{nm}$, these
wave packets pass all fundamental physical requirements.
The first three normalized wave packets on the list are given
explicitly by
\begin{eqnarray}
	&&\psi_0=\frac{\sin^{\frac{1}{4}} 2\varphi
	~e^{-\frac{x^2}{2~}tan(\varphi-i\rho)}}{(2\pi)^{\frac{1}{4}}
	\cosh^{\frac{1}{2}} (\rho+i\varphi) }~, \\
	&&\psi_1=\frac{\sin^{\frac{3}{4}} 2\varphi
	~x e^{-\frac{x^2}{2~}tan(\varphi-i\rho)}}{(2\pi)^{\frac{1}{4}}
	\cosh^{\frac{3}{2}} (\rho+i\varphi) }~, \\
	&&\psi_2=\frac{\sin^{\frac{5}{4}} 2\varphi
	~(x^2-\delta^2 (\rho,\varphi))
	e^{-\frac{x^2}{2~}tan(\varphi-i\rho)}}{\sqrt{2}(2\pi)^{\frac{1}{4}}
	\cosh^{\frac{5}{2}} (\rho+i\varphi) } ~.
\end{eqnarray}
Notice that, reflecting their non-trivial $\rho$-dependence, the
polynomials involved are not the Hermite polynomials.
The ground state $\psi_0$ has been introduced by Guth-Pi and
Barton \cite{GP}, with $t$ replacing $\rho$ of course, when
discussing the quantum mechanics of the scalar field in the
so-called new inflationary universe.
The raising and lowering operators are given by
\begin{equation}
	b^{\pm}=\frac{\pm i}{\sqrt{\sin 2 \varphi}}
	\left(p\cosh (\rho\mp i\varphi)
	-x \sinh(\rho\mp i\varphi)\right)
\end{equation}
\begin{eqnarray}
	&& b^- \psi_n=(-1)^n\sqrt{n}~\psi_{n-1} ~,\\ 
	&& b^+ \psi_n=(-1)^{n+1}\sqrt{n+1}~\psi_{n+1}~,
\end{eqnarray}
giving the Hamiltonian the form
\begin{equation}
	{\cal H}=-\frac{{b^+}^2 +{b^-}^2
	+\cos 2\varphi ~(b^+ b^- +b^- b^+)}{2\sin 2\phi} ~.
	\label{Hbb}
\end{equation}
Owing to $\langle {b^\pm}^2 \rangle=0$,
associated with the wave packets are then the global, meaning
$\rho$-independent, average-'energy' levels
\begin{equation}
	E_n=\int_{-\infty}^{\infty} \psi^{\dagger}_n {\cal H}
	\psi_n ~dx=-\left(n+ \frac{1}{2}\right)\cot{2\varphi} ~.
	\label{E0}
\end{equation}
The choice $\cot{2\varphi}<0$ gives rise to a positive spectrum, and
combining with the previous integrability condition $\sin{2\varphi}>0$,
the still arbitrary angle $\varphi$ gets restricted to the region
$\frac{\pi}{4}<\varphi<\frac{\pi}{2}$.

Constructing the set of localized wave packets, we can now
calculate quantum mechanical expectation values associated with
the various metric component operators.
To do so, we  first notice a direct consequence of the discrete
symmetry
$x \rightarrow -x$, namely $\langle x \rangle_n=\langle p \rangle_n=0$.
And then, recalling the relations eq.(\ref{xp}), we find
\begin{eqnarray}
	&\langle 1-R \rangle_n=0 ~,~
	\langle (1-R)^2 \rangle_n
	=\displaystyle{\frac{(2n+1)e^{-2\rho}}{2\sin 2\varphi} ~, }&\label{msquare}\\
	&\langle p_R \rangle_n=0 ~,~
	\displaystyle{\langle p_R^2\rangle_n=\frac{(2n+1)
	e^{2\rho}}{2\sin 2\varphi}~,}& \label{pRsquare}
\end{eqnarray}
with the associated uncertainty relation reading
\begin{equation}
	\Delta R ~\Delta p_R
	=\frac{2n+1}{2\sin 2\varphi} \geq \frac{1}{2}~,
	\label{uncertain}
\end{equation}
in accord with the Dirac brackets eq.(\ref{RT}).
These formulas allow us to make contact with the familiar
general relativistic Schwarzschild solution $1-R=2m/r$
and $T/R=p_R^2/r^2 =const$
(the constant can always be absorbed by a time rescaling),
with the dictionary reading $r=e^\rho$.
We refer to the emerging spacetime as the quantum mechanical
Schwarzschild vacuum.
It is massless on average, but exhibits a non-zero mass
uncertainty
\begin{equation}
	\langle m \rangle_n \pm \Delta m_n
	=0 \pm \sqrt{\frac{2n+1}{8\sin 2\varphi}}~,
\end{equation}
which can be interpreted an equal amount of positive and
negative mass metric fluctuations.
Note that the quantum uncertainty is bounded from below
$\Delta m\geq \frac{1}{2\sqrt{2}}$ and cannot disappear.
%%%%%
In other words, Minkowski spacetime is only flat in average.
%%%%%

One may wonder though what goes wrong
when attempting to construct an eigenvacuum $\Psi_0$,
namely a zero eigenmass state of the
mass operator
$\hat{m}\sim e^{i\varphi}b^- +e^{-i\varphi}b^+$
(see the forthcoming eq.(\ref{hatm})).
It takes some algebra to prove that
$\Psi_0 \sim \sum_{k} c_k\psi_{2k}$,
with the coefficients subject to the series expansion
\begin{equation}
	(1-y^2)^{-1/2}=\sum_{k} |c_k|^2 y^{2k}~.
	\label{div}
\end{equation}
The sum $\sum_k|c_k|^2$ diverges, giving rise to
unacceptable quantum mechanical  consequences.

\medskip
\noindent {\underline{\bf{The massive case}}

The inclusion of mass requires the violation of the discrete
$x\rightarrow -x$ symmetry of the vacuum wave function.
This is done by simply shifting the Gaussian of the Guth-Pi-Barton
tower, with the elaborated wave functions taking the form
\begin{equation}
	\psi_n (x,\rho)
	=\tilde{P}_n (x,\rho)
	 ~e^{-\frac{x^2}{2~}tan(\varphi-i\rho)
	 -\eta x \sec (\varphi-i \rho)} ~,
	 \label{Pe}
\end{equation}
introducing the shift parameter $\eta$ designated to induce the
mean Schwarzschild mass.
The modified polynomials $\tilde{P}_n (x,\rho)$ generalize the previous
$P_n (x,\rho)$, and contain now even as well as odd powers of $x$.
The first normalized wave packets are given explicitly by
\begin{eqnarray}
	&& \psi_0=\frac{\sin^{\frac{1}{4} }2\varphi~
	e^{-\frac{1}{2}(x^2+\eta^2) tan(\varphi-i\rho)
	 -\eta x \sec (\varphi-i \rho)}}
	{(2\pi)^{\frac{1}{4}}e^{\frac{1}{2}\eta^2 \cot \varphi}
	\cosh^{\frac{1}{2}(\rho+i\varphi})} ~, \\ \nonumber
	&& \psi_1=\left(\eta \cot \phi+\frac{x-i \eta \sinh(\rho
	+i\varphi)}{\cosh (\rho+i\varphi)}\right) \\ 
	&& \quad \quad \frac{\sin^{\frac{3}{4} }2\varphi
	~e^{-\frac{1}{2}(x^2+\eta^2) tan(\varphi-i\rho)
	 -\eta x \sec (\varphi-i \rho)}}
	{{(2\pi)^{\frac{1}{4}}e^{\frac{1}{2}\eta^2 \cot \varphi}
	\cosh^{\frac{1}{2}(\rho+i\varphi})}}~.
\end{eqnarray}
The raising and lowering operators get shifted
\begin{equation}
	b^{\pm}_{\eta}=b^{\pm}
	-\frac{\eta}{\sqrt{\sin 2\varphi}}~,
\end{equation}
in obvious notations. 
The Hamiltonian in the new basis resembles eq.(\ref{Hbb}),
with $b_\eta^\pm$ replacing $b^\pm$,
but gets further supplemented by
\begin{equation}
	-\frac{\eta \cot \varphi}{\sqrt{\sin 2\varphi}}
	(b_{\eta}^+ + b_{\eta}^-)-\frac{\eta^2}{2 \sin ^2 \varphi}
\end{equation}
Owing to $\langle b_\eta^\pm \rangle=0$,
associated with the new set is again a
ladder average-'energy' spectrum, but it is now uniformly
shifted relative to the vacuum ladder.
To be specific,
\begin{equation}
	E_n=-\left( n+\frac{1}{2}\right)\cot 2\varphi
	-\frac{\eta^2}{2\sin^2 \varphi} ~.
	\label{shift}
\end{equation}
Had we adopted the special values
$\eta^2_\ell=\ell (\tan\varphi-\sin 2\varphi)$ ($\ell$ integer)
we would have in fact recaptured the vacuum average-'energy' ladder
\begin{equation}
	E_{n,\ell}=
	-\left(n-\ell+ \frac{1}{2}\right)\cot{2\varphi} ~.
	\label{Eell}
\end{equation}
The various massive wave packets are characterized by the
$n$-independent quantum averages
\begin{equation}
	\langle x \rangle_n=
	-\frac{\eta \cosh \rho}{\sin \varphi} ~,\quad
	\langle p \rangle_n=
	-\frac{\eta \sinh \rho}{\sin \varphi}  ~,
\end{equation}
and hence share the one and the same classical Schwarzschild
metric, with
$\frac{1}{\sqrt{2}}\langle p-x \rangle=2\langle m \rangle /r$.
The mass operator itself can be expressed in terms of the raising
and lowering operators
\begin{equation}
	\hat{m}=\frac{1}{2\sqrt{2}}\left(
	\frac{\eta}{\sin \varphi}
	+\frac{ e^{i\varphi}b^-+e^{-i\phi}b^
	+}{\sqrt{\sin 2\varphi}}\right)~.
	\label{hatm}
\end{equation}
The two uncertainties $\Delta R$ and $\Delta p_R$ turn
out to be insensitive to the presence of the $\eta$-parameter, 
retaining the exact vacuum value, with eq.(\ref{uncertain})
untouched.
Altogether, associated with the quantum mechanical Schwarzschild
black hole of the $n$-th state is the mass formula
\begin{equation}
	\langle m \rangle_n \pm \Delta m_n
	=\frac{\eta }{2\sqrt{2}\sin \varphi}
	\pm \sqrt{\frac{2n+1}{8\sin 2\varphi}}~.
	\label{mass}
\end{equation}
Several remarks are in order:

\noindent (i) 
The sign of $\eta$ is as arbitrary as the sign of the mass parameter
in the original Schwarzschild solution.

\noindent (ii)
The larger $\eta$ is, the more negligible is the $\frac{\Delta m}{m}$
ratio, driving the solution into a more classical regime.

\noindent (iii)
The larger is $\eta$, the larger is the uniform shift downwards,
see eq.(\ref{shift}), of the average-'every' levels.
This in turn increases the number of the low 'energy' states which
actually penetrate the upside-down harmonic potential barrier.

\noindent (iv)
While  the underlying classical gravitational metric is governed
by $\langle 1-R \rangle$, it is independent of
$\langle p_R \rangle$ which can be absorbed by a time
rescaling.

%%%%%
\noindent (v)
Note that the introduction of the $\eta$ parameter does not
resolve the divergence problem encountered in eq.(\ref{div}).
Exactly like in the massless case, reflecting the fact that the
eigenstate $\Psi_{\eta}$ of eigenmass proportional to $\eta$
is not square integrable, it cannot be expressed as a linear
combination of the wave packets $\psi_n$.
%%%%%

The wave functions $\psi_n$ are not sharply peaked about
any particular classical trajectory, and in particular, do not
seem to exhibit any exceptional behavior at
$r=2\langle m \rangle$ which marks the classical location
of the black hole event horizon.
However, the fact that the universal variance eq.(\ref{msquare})
is kept unchanged in the massive sector may have important
consequences for black hole thermodynamics.
The central geometrical role here is played by the horizon surface
area $A$.
Classically, we know that $A=16\pi m^2$ for $m>0$, but this
leaves the door quantum mechanically open for the ambiguity
$\langle A\rangle \sim \langle m^2 \rangle$ versus
$\langle A \rangle\sim \langle m \rangle^2$.
To make a decision, we remind the reader that treating the
horizon surface area as an adiabatic invariant, an equally
spaced Bohr-Sommerfeld area spectrum has been conjectured
by Bekenstein \cite{B} and subsequently modelled by
Bekenstein-Mukhanov \cite{BM}.
In our case, while $\langle m \rangle$ in $n$-independent,
it is $\langle m^2 \rangle$ which is linearly quantized as
required, implying
\begin{equation}
	\langle A \rangle_n=16\pi \langle m^2\rangle_n
	=2\pi\left(\frac{\eta^2}{\sin^2 \varphi}
	+ \frac{2n+1}{\sin 2\varphi}\right)~.
	\label{A}
\end{equation}
Representing the vacuum structure, the existence of a minimal
surface area $\langle A \rangle_{min}=2\pi/{\sin 2\varphi}$
is noticeable, advocating the case of a universally fixed $\varphi$.
It should be emphasized, however,  that despite of the
apparent similarity, eq.(\ref{A}) differs from
Bekenstein quantization.
While any two distinct states labeled by $n_{1}\neq n_{2}$ are
conventionally interpreted to be associated with two distinct
black holes of masses $m_1\neq m_2$, they are associated
in our case with a common  $\langle m \rangle$.
A closer inspection reveals that Bekenstein's conjecture can be
anchored to eq.(\ref{Eell}), with the minimal massive horizon
surface area is then
$\langle A\rangle_{\ell=1,n=0} =2\pi \tan\varphi$.

The discussion presented in this paper, while hopefully
shedding some light on what to expect when letting
quantum mechanics meet general relativity, leaves a bunch
of question marks open.
In particular, like in previous approaches, the statistical role
of the average-'energy' ladder
$E_n$ is yet to be challenged by black hole thermodynamics.
Also, $\varphi$ is an arbitrary parameter at this stage, but
there is a good reason to suspect that it is uniquely fixed.
At any rate, looking at the half full glass, we have demonstrated
that one can
(i) Probe the vacuum/massive Schwarzschild black hole
quantum mechanics even though quantum gravity is still
absent,
(ii) Reveal the universal quantum mechanical structure of
the Schwarzschild black hole geometry, and
(iii) Do it without appealing to theories beyond general relativity,
such as string theory or loop gravity.
For a sequel of this paper, introducing the idea of thermal
Hawking broadening of black hole wave packets, see ref.(\cite{thermal}) .

\acknowledgments
{We are very thankful to Ilya Gurwich for a constructive discussion,
and especially for asking the right question at the right time.
Special thanks to BGU president Prof. Rivka Carmi for the
kind support.}

\end{document}